\documentclass[lettersize,journal,onecolumn,draftclsnofoot,12pt]{IEEEtran}
\usepackage{amsmath,amsfonts}
\usepackage{algorithm}
\usepackage{algpseudocode}  
\usepackage{array}
\usepackage[caption=false,font=normalsize,labelfont=sf,textfont=sf]{subfig}
\usepackage{textcomp}
\usepackage{stfloats}
\usepackage{url}
\usepackage{verbatim}
\usepackage{graphicx}
\usepackage{cite}
\usepackage{multicol}
\usepackage{ragged2e} 
\usepackage{booktabs}
\usepackage{float}
\usepackage{afterpage}
\usepackage{tikz}
\usepackage{pgfplots}
\pgfplotsset{compat=1.18}
\DeclareUnicodeCharacter{200E}{}
\DeclareUnicodeCharacter{200B}{}
\DeclareUnicodeCharacter{200F}{}
\begin{document}

\title{Optimized Beamforming and Bandwidth Allocation in Multi-Antenna UAV-Assisted Vehicular Networks}

\author{S.  Fatemeh Bozorgi\thanks{S. F. Bozorgi and S. M. Razavizadeh are with the School of Electrical Engineering, Iran University of Science \& Technology  (IUST), Tehran, Iran (e-mail: sf\textunderscore bozorgi‎@elec.iust.ac.ir; smrazavi@iust.ac.ir).}, S. Mohammad Razavizadeh,~\IEEEmembership{Senior Member, IEEE,} and Jiguang He,~\IEEEmembership{Senior~Member,~IEEE}\thanks{J. He is with the School of Computing and Information Technology, Great Bay University, Dongguan 523000, China and Great Bay Institute for Advanced Study (GBIAS), Dongguan 523000, China (e-mail: jiguang.he@gbu.edu.cn).}} 
     





\maketitle

\begin{abstract}
Ensuring reliable communication for mission-critical vehicles in dynamic environments with limited infrastructure is a significant challenge due to interference and spectrum scarcity. This paper investigates a UAV-assisted vehicular communication framework that leverages multi-antenna beamforming and dynamic bandwidth allocation to provide prioritized and interference-mitigated wireless links. Vehicles are classified according to their service priority, with each class assigned a distinct frequency band to reduce interference. Within each class, optimized beamforming further minimizes transmission overlap and enhances spectral efficiency. 
The optimization problem is solved using an alternating optimization framework, incorporating two beamforming strategies: one based on successive convex approximation (SCA) and the other derived in closed form. Numerical results indicate that the proposed scheme outperforms baseline approaches that optimize only bandwidth allocation or beamforming in terms of overall system performance. Among the two joint optimization methods, the closed-form solution achieves higher sum rates and generally requires less transmit power, while also exhibiting lower computational complexity compared with the SCA-based approach.

\end{abstract}


\section{Introduction}
Emergency vehicles-such as ambulances-are frequently deployed in situations that demand prompt and reliable information exchange to support their operational effectiveness. In environments such as traffic accidents or disaster-affected zones, maintaining stable communication links plays a critical role in enhancing coordination and facilitating situational awareness [1].

To address these communication demands, vehicle-to-everything (V2X) technologies provide a comprehensive set of communication modes-including vehicle-to-vehicle (V2V), vehicle-to-infrastructure (V2I), vehicle-to-pedestrian (V2P), and vehicle-to-network (V2N)-enabling efficient information exchange among vehicles, infrastructure, pedestrians, and other network participants, thereby enhancing traffic safety and operational efficiency. As an example, V2I links enable the delivery of timely information regarding traffic conditions and roadway status, assisting vehicles in making context-aware navigation decisions. Employing these communication modalities can enhance coordination among various vehicle categories and support sustained network reliability across diverse traffic scenarios and connectivity demands [2].

Additionally, unmanned aerial vehicles (UAVs) have demonstrated strong potential in enhancing wireless communication systems, particularly in scenarios where conventional infrastructure is either absent or impaired. Due to their rapid deployment capabilities and elevated operating positions, UAVs can offer flexible and wide-area wireless coverage, contributing to the continuity and resilience of communication links in demanding environments [3]. A notable benefit of UAV-assisted communication is the ability to establish and maintain favorable line-of-sight (LoS) channels with ground users in many scenarios, thereby enhancing signal strength and reducing transmission delays [4]–[5]. UAVs can also be utilized as aerial base stations, relays, or data collection units, enabling their application across a variety of wireless networking scenarios. 

Accordingly, several studies have investigated resource allocation and interference management in UAV-assisted wireless networks. In [6], joint user association and subchannel allocation with spectrum reuse are studied to enhance the achievable data rates of ground users. In [7], users are divided into delay-sensitive and delay-tolerant groups, and joint bandwidth and transmit power allocation is employed to guarantee the rate requirements of delay-sensitive users while maximizing the minimum achievable rate of the remaining users. Moreover, UAV-assisted vehicular networks in cooperation with terrestrial infrastructure are considered in [8], where quality-of-experience improvement is addressed under interference and backhaul constraints.

Although several existing studies still consider single-antenna UAVs or simplified network configurations, equipping UAVs with multiple antennas enables advanced beamforming techniques that can significantly enhance spectral efficiency, increase achievable data rates, and support simultaneous transmission to multiple users~[9]. In particular, multi-antenna UAVs can exploit beamforming to achieve constructive signal combination at the receiver, thereby improving the effective signal strength in multi-user scenarios.

In this context, [10] investigates a multi-user UAV-assisted MISO communication system and optimizes transmit beamforming to enhance the achievable sum-rate, demonstrating the throughput gains enabled by multi-antenna UAV transmission. In [11], a UAV-enabled multi-user system equipped with a multi-antenna array is studied from an energy transmission perspective, where analog beamforming is employed to improve spatial energy concentration, highlighting the role of antenna array configuration and beam control.
Furthermore, [12] considers multi-UAV networks and proposes a distributed beamforming strategy in which multiple UAVs cooperatively transmit to improve the achievable sum-rate under per-antenna power constraints. [13] investigates joint UAV deployment and transmit beamforming with the objective of maximizing the minimum communication rate under beam pattern and power constraints.
In addition, [14] studied an integrated satellite and aerial network where a multi-antenna UAV serves multiple users, and proposed a joint beamforming and resource allocation framework to maximize the achievable sum rate while ensuring computation accuracy.

Despite the significant progress in UAV-assisted wireless networks including the above studies, their application in vehicular communication systems from the perspective of joint resource allocation has not yet been adequately investigated.

In this study, we consider a vehicular communication scenario in which vehicles are categorized into two groups: emergency and regular. This classification is particularly important in situations involving emergency vehicles, which typically travel at higher speeds and generally require more stable connectivity to fulfill their critical roles effectively [15]. In this context, our previous work in [16] employed a single-antenna UAV with joint trajectory and bandwidth optimization to serve vehicular users. Building upon that concept, the present study addresses other challenges, such as interference management and resource coupling, that arise in multi-UAV multi-antenna systems.

Key contributions are summarized as follows:\\
\begin{itemize}
\item We consider a priority-based service model that classifies vehicles into emergency and regular categories, where each emergency vehicle must satisfy a minimum rate constraint while the sum rate of regular vehicles is maximized. 
\item We adopt an orthogonal frequency subband allocation between the two vehicle groups to completely eliminate inter-group interference, while allowing bandwidth reuse within each group to efficiently handle overlapping content (e.g., common map updates or hazard alerts).
\item We develop a distributed beamforming framework using a network of multi-antenna UAVs to mitigate intra-group interference, and we design two distinct beamforming strategies: an iterative SCA-based method and a closed-form solution.
\item We show that the closed-form solution achieves higher sum rates and lower power consumption than the SCA-based approach, together with lower computational complexity.
\item Furthermore, we develop a joint optimization framework that integrates bandwidth allocation and distributed beamforming design. This framework aims to maximize the sum data rate of regular vehicles, thereby improving overall communication efficiency in the network.
\end{itemize} 

To the best of our knowledge, despite the importance of jointly managing interference, power, and bandwidth in UAV-enabled wireless systems, the specific combination of system model and optimization formulation considered in this work—integrating multi-UAV multi-antenna beamforming with bandwidth allocation under quality of service (QoS) requirements—has not been addressed in the existing literature.

The remainder of this paper is organized as follows. Section II provides a detailed description of the system model, along with the mathematical formulation of the optimization problem. In Section III, two iterative solution approaches are presented: SCA-based algorithm and a closed-form method. Section IV presents the simulation results, highlighting the performance and resource efficiency of the proposed scheme. Finally, Section V concludes the paper with a summary of the main findings.

\emph{Notations}:
In this paper, scalar quantities are denoted by italic letters, vectors by bold lowercase letters, and matrices by bold uppercase letters. The Euclidean norm of a vector and the Frobenius norm of a matrix are both denoted by $\|\cdot\|$, depending on the context. The transpose, complex conjugate, and Hermitian (conjugate transpose) operations are represented by $(\cdot)^T$, $(\cdot)^*$, and $(\cdot)^H$, respectively.
The real part and modulus of a complex number are denoted by $\Re\{\cdot\}$ and $|\cdot|$, respectively. The trace of a matrix is indicated by $\mathrm{Tr}(\cdot)$, and the $(i,j)$-th entry of a matrix is written as $(\cdot)_{i,j}$. In addition, the base-2 logarithm is represented by $\log_2(\cdot)$, and the imaginary unit is defined as $j = \sqrt{-1}$. The identity matrix is denoted by $\mathbf{I}$, and optimal variable values are indicated with a superscript $\star$, such as $x^\star$.

\section{System Model}
In this study, we consider the deployment of a group of $M$ UAVs, each equipped with a uniform linear array (ULA) of $L$ antennas, to provide wireless communication services to $V$ ground vehicles. The UAVs utilize distributed beamforming techniques to ensure reliable and efficient data transmission.

Fig. 1 illustrates the system model, where UAVs are deployed to cover a road area with dimensions $E_x$ and $E_y$, enabling seamless wireless connectivity across the region.

The primary information about the vehicles, such as their locations and speeds, is made available to the UAVs [17]. Although the system is modeled over a short time interval under quasi-static assumptions for simplicity, the framework can be extended to dynamic scenarios by sequentially solving the optimization over consecutive time blocks. In each block, resource allocation is based on slowly varying large-scale channel state information (CSI), which remains approximately constant due to location-dependent variations [18]. Between blocks, vehicle and UAV positions are updated, and the optimization is re-executed with the new large-scale CSI. Vehicles that exit the coverage area of the current UAVs can be reassigned to neighboring UAVs or trigger a handoff, without altering the core alternating optimization structure.
Furthermore, the instantaneous position of the $m$th UAV can be represented as $(x_m, y_m, z_m)$. Consequently, the distance between each UAV and every vehicle is calculated as follows:

\begin{figure}[]
\begin{center}
\centering
\includegraphics[trim={0cm 0cm 0cm 0cm},clip, width=8.5cm, height=5cm]{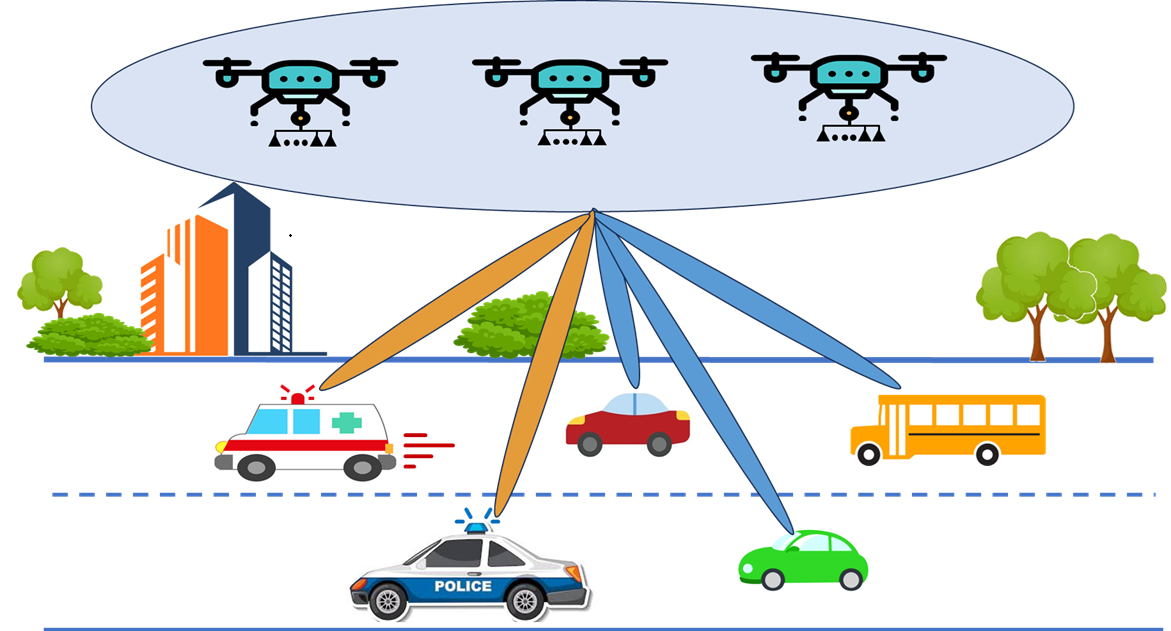}
\caption{System model of distributed beamforming in vehicular networks. Emergency vehicles are provided with a data rate above a predefined threshold.}
\label{fig1}
\end{center}
\end{figure}

\begin{equation}
\begin{aligned}
d_v^m &= \sqrt{(x_v - x_m)^2 + (y_v - y_m)^2 + z_m^2} \\
      &= \sqrt{\| \mathbf{p}_v - \mathbf{p}_m \|^2 + z_m^2},
\end{aligned}
\end{equation}
where ${{{\bf{p}}_v} = [{x_v},{y_v}]}$ and ${{{\bf{p}}_m} = [{x_m},{y_m}]}$ represent the positions of the vehicles and the projections of the UAVs' positions onto the ground, respectively.

The wireless channel between the $v$th vehicle and all UAVs, which is considered to be LoS, is represented by a concatenated vector ${\mathbf{h}}_v^{ML \times 1} = {[{\mathbf{h}}_v^1,...,{\mathbf{h}}_v^m,...,{\mathbf{h}}_v^M]^T}$, where ${\mathbf{h}}_v^m$ is given by

\begin{equation}
\begin{gathered}
  {\mathbf{h}}_v^m = \frac{{{d_0}}}{{d_v^m}} \left[1,\; e^{j\pi \cos {\varphi _m}},\; \ldots,\; e^{j\pi (L - 1)\cos {\varphi _m}} \right], \hfill \\
  \cos {\varphi _m} = \frac{{x_m - x_v}}{{d_v^m}}, \hfill \\ 
\end{gathered} 
\end{equation}
where $d_0$ is the average channel gain at a reference distance of 1 meter, and ${\varphi _m}$ denotes the angle between the LoS vector from the UAV to the vehicle and the $x$-axis, along which the ULA is placed.
It is worth noting that, according to 3GPP Release 15, the LoS probability increases significantly with UAV altitude [19]-[20]. Given our considered altitude and following the common practice in many prior works [4], [7], [10]--[11], [13], [16]--[17], [25], we adopt LoS model here.

In the described UAV-assisted vehicular communication system, a total of $V$ ground vehicles are present, among which a subset of $V_1$ vehicles---identified by $\Phi_1 = \{1, 2, \ldots, V_1\}$---operate as emergency units, including ambulances, fire trucks, and police vehicles dispatched in response to natural disasters or large-scale emergencies. To ensure timely access to critical information---such as real-time traffic updates, road conditions, and coordination data---each of these vehicles must maintain a data rate above a specified threshold. This requirement is crucial for enabling uninterrupted communication during operations, where delays or packet losses could hinder emergency response efforts and compromise safety.
For the remaining vehicles, which are identified by the set ${v_2} \in { \Phi _2}=\{ 1,2,...,V - {V_1}\} $, the objective is to maximize the sum of their average achievable data rates across the network. The instantaneous rate expressions for both regular and emergency vehicles are defined as follows:

\begin{equation}
\begin{gathered}
  R_{v_1} = B b_1 \log_2 \left( 1 + \frac{|\mathbf{h}_{v_1}^H \mathbf{w}_{v_1}|^2}{\sum\limits_{i = 1, i \ne v_1}^{V_1} |\mathbf{h}_{v_1}^H \mathbf{w}_i|^2 + \sigma^2} \right), \hfill \\
  R_{v_2} = B b_2 \log_2 \left( 1 + \frac{|\mathbf{h}_{v_2}^H \mathbf{w}_{v_2}|^2}{\sum\limits_{i = 1, i \ne v_2}^{V_2} |\mathbf{h}_{v_2}^H \mathbf{w}_i|^2 + \sigma^2} \right), \hfill
\end{gathered}
\end{equation}
where,${{\mathbf{w}}_{v_l}} \in {\mathbb{C}^{ML \times 1}},{v_l} \in {\Phi _l},l \in \{ 1,2\} $ denotes the beamforming vector assigned to the $v$th vehicle, and ${{\sigma ^2}}$ represents the power of the additive white Gaussian noise (AWGN) at the receiver. The system adopts an orthogonal frequency-division multiple access (OFDMA) scheme to serve the two groups of vehicles---emergency and regular---on orthogonal frequency subbands, thereby preventing inter-group interference. However, within each group, the allocated bandwidth is reused among the corresponding vehicles, since their transmitted content may partially overlap—for instance, in the form of group-level map updates or common alert messages. This reuse gives rise to intra-group interference, as captured in the denominator of the rate expression in (3). Moreover, the total system bandwidth is denoted by $B$. The bandwidth allocation coefficients $b_l$, $l \in \{1,2\}$, specify the portion of $B$ assigned to vehicle group $l$, where group~1 corresponds to emergency vehicles and group~2 to regular vehicles. The allocation factors ${b_l}$​ are assumed to be known at each time slot via control signaling and are subject to the following constraints:

\begin{equation}
\begin{array}{l}
0 \le b_l \le 1, \quad l \in \{1,2\} \\
 b_1+b_2 \le 1.
\end{array}
\end{equation}

Based on the above considerations, the optimization problem is formulated as follows:
\begin{subequations}\label{eq:2}
\begin{align}
&\mathop{\max}\limits_{\substack{{\mathbf{w}}_{v_l},\, l \in \{1,2\} \\ b_1,\, b_2}} 
\sum\limits_{v_{2} = 1}^{V_2} B b_2 \log_2 \left( 1 + \frac{|\mathbf{h}_{v_2}^H \mathbf{w}_{v_2}|^2}{\sum\limits_{i = 1, i \ne v_2}^{V_2} |\mathbf{h}_{v_2}^H \mathbf{w}_i|^2 + \sigma^2} \right) \\
\text{s.t.} \quad
& B b_1 \log_2 \left( 1 + \frac{|\mathbf{h}_{v_1}^H \mathbf{w}_{v_1}|^2}{\sum\limits_{i = 1, i \ne v_1}^{V_1} |\mathbf{h}_{v_1}^H \mathbf{w}_i|^2 + \sigma^2} \right) \geq R_{\text{th}}, \\
& \sum_{v=1}^{V} \big\| \mathbf{w}_{v,m} \big\|^2 \;\le\; P_m, \qquad \forall m = 1,\dots,M, \\
& 0 \le b_l \le 1, \quad l \in \{1,2\}, \\
& b_1 + b_2 \le 1.
\end{align}
\end{subequations}

Constraint (5b) specifies that each emergency vehicle must achieve a data rate above the threshold $R_{th}$. Meanwhile, constraint (5c) enforces an individual power budget for each UAV, where $\mathbf{w}_{v,m} \in \mathbb{C}^{L \times 1}$ denoting the beamforming vector of UAV $m$ for vehicle $v$, and $P_m$ is the maximum transmit power of UAV $m$.

The above problem is non-convex due to the interference-coupled rate expressions and the joint optimization over beamforming vectors and bandwidth allocation. To address this, we adopt an alternating optimization strategy, where each subset of variables is updated iteratively while keeping the others fixed, leading to a sequence of more tractable subproblems.

\section{Proposed Method}
To handle the non-convex nature of the main optimization problem in (5) and to enable efficient solution development, we employ an alternating optimization approach that decomposes the problem into two more tractable sub-problems. In the first sub-problem, the bandwidth allocation is optimized while the beamforming vectors are held fixed. Then, based on the obtained bandwidth allocation, the second sub-problem focuses on optimizing the beamforming design. The following subsections provide a detailed treatment of both sub-problems.
 
\subsection{Bandwidth Optimization}
In this subsection, we aim to determine an efficient allocation of the available spectral resources between the two vehicle groups to enhance overall system performance. Given that the system operates under an OFDMA scheme, emergency and regular vehicles are assigned orthogonal frequency subbands, inherently eliminating inter-group interference. Accordingly, this section is devoted to optimizing the bandwidth allocation coefficients ${b_l},l \in \{ 1,2\}$, as defined in Problem (6):

\begin{subequations}\label{eq:2}
\begin{align}
&\mathop{\max}\limits_{b_1, b_2} \sum\limits_{v_{2} = 1}^{V_2} B b_2 \log_2 \left( 1 + \frac{|\mathbf{h}_{v_2}^H \mathbf{w}_{v_2}|^2}{\sum\limits_{i = 1, i \ne v_2}^{V_2} |\mathbf{h}_{v_2}^H \mathbf{w}_i|^2 + \sigma^2} \right)\\
\text{s.t. } \quad
& B b_1 \log_2 \left( 1 + \frac{|\mathbf{h}_{v_1}^H \mathbf{w}_{v_1}|^2}{\sum\limits_{i = 1, i \ne v_1}^{V_1} |\mathbf{h}_{v_1}^H \mathbf{w}_i|^2 + \sigma^2} \right) \geq R_{\text{th}}, \\
& 0 \le b_l \le 1, \quad l \in \{1,2\}, \\
& b_1 + b_2 \le 1.
\end{align}
\end{subequations}

In Problem (6), given that the beamforming vectors are fixed, both the objective function and the constraints are linear with respect to the bandwidth allocation coefficients. Consequently, the problem is convex and amenable to solution via standard convex optimization methods, such as the interior-point algorithm [21].

\subsection{Beamforming Optimization}

Once the bandwidth allocation has been determined, the subproblem proceeds with optimizing the beamforming design for multiple multi-antenna UAVs. The corresponding optimization, denoted as subproblem (7), is formulated as:
\begin{subequations}\label{eq:2}
\begin{align}
\max_{\mathbf{w}_{v_l},\, l \in \{1,2\}} \quad
& \sum_{v_2 = 1}^{V_2} B b_2 
\log_2 \!\left(
1 +
\frac{|\mathbf{h}_{v_2}^H \mathbf{w}_{v_2}|^2}
{\sum_{i = 1, i \ne v_2}^{V_2}
|\mathbf{h}_{v_2}^H \mathbf{w}_i|^2 + \sigma^2}
\right) \\[1mm]
\text{s.t.} \quad
& B b_1 \log_2 \!\left(
1 +
\frac{|\mathbf{h}_{v_1}^H \mathbf{w}_{v_1}|^2}
{\sum_{i = 1, i \ne v_1}^{V_1}
|\mathbf{h}_{v_1}^H \mathbf{w}_i|^2 + \sigma^2}
\right)
\ge R_{\text{th}}, \\
& \sum_{v=1}^{V} \big\| \mathbf{w}_{v,m} \big\|^2 \le P_m,
\qquad \forall m = 1,\dots,M .
\end{align}
\end{subequations}

 To address subproblem (7), two different approaches are proposed: the first utilizes the SCA method to iteratively solve the non-convex problem, while the second offers a closed-form solution. Detailed formulations and algorithmic procedures for both methods are presented in the subsequent subsections.

\subsubsection{Beamforming Design via the SCA Method}
To facilitate the beamforming vector design, we introduce the lifted covariance matrices 
${\mathbf{C}}_{v} = {\mathbf{w}}_{v}{\mathbf{w}}_{v}^H$ for $v \in \{1,\ldots,V\}$, 
and denote by ${\mathbf{C}}_{v_l}$, $v_l \in \Phi_l$, the matrices associated with vehicles in class $l \in \{1,2\}$. 
Based on this transformation and by applying the semidefinite relaxation (SDR) technique, 
the beamforming optimization subproblem can be reformulated as follows:
\begin{subequations}\label{eq:2}
\begin{align}
& \mathop{\max}_{{\mathbf{C}}_{v_l},\, l \in \{ 1,2\}} 
\sum_{v_2 = 1}^{V_2} B b_2 \log_2\left(1 + 
\frac{\operatorname{Tr}({\mathbf{A}}_{v_2} {\mathbf{C}}_{v_2})}
{\sum\limits_{\substack{i = 1 \\ i \ne v_2}}^{V_2} \operatorname{Tr}({\mathbf{A}}_{v_2} {\mathbf{C}}_{i}) + \sigma^2} \right) \\
\text{s.t.} \quad
& B b_1 \log_2\left(1 + 
\frac{\operatorname{Tr}({\mathbf{A}}_{v_1} {\mathbf{C}}_{v_1})}
{\sum\limits_{\substack{i = 1 \\ i \ne v_1}}^{V_1} \operatorname{Tr}({\mathbf{A}}_{v_1} {\mathbf{C}}_{i}) + \sigma^2} \right)
\geq R_{\text{th}}, \\
& {\mathbf{C}}_{v_l} \succeq 0,\quad \operatorname{rank}({\mathbf{C}}_{v_l}) = 1,\quad v_l \in \Phi _l,\quad l \in \{1,2\}, \\
& \sum_{v=1}^{V} \operatorname{Tr}({\mathbf{E}}_{{m}}{\mathbf{C}}_{{v}}) \leq P_m, \qquad \forall m = 1,\dots,M, 
\end{align}
\end{subequations}
where ${\mathbf{A}}_{v_l} = {\mathbf{h}}_{v_l} {\mathbf{h}}_{v_l}^H$ and ${\mathbf{E}}_m \in \mathbb{R}^{ML \times ML}$ denotes the block-diagonal selection matrix for UAV $m$, whose $m$-th diagonal block is ${\mathbf{I}}_{L}$ and all other entries are zero. The matrices ${\mathbf{C}}_{v_l}$ are required to be positive semidefinite and rank-one to correspond to valid beamforming vectors. However, the rank-one constraint is non-convex and complicates direct solution. To simplify the problem, this constraint is relaxed, resulting in a convex semidefinite program (SDP) solvable via standard convex optimization tools. If the relaxed solution is rank-one, the beamforming vector can be recovered from its principal eigenvector; otherwise, approximation methods such as Gaussian randomization are applied.

Accordingly, by applying the SCA method\footnotemark, problem (8) is reformulated as follows:

\footnotetext{The SCA method addresses non-convex problems by iteratively constructing and solving convex surrogate subproblems that approximate the original objective and constraints around the current iterate [22].}

\begin{subequations}\label{eq:2}
\begin{align}
& \mathop {\max }\limits_{{{{\mathbf{C}}_{{v_l}}},l \in \{ 1,2\} }} \sum_{{v_2} = 1}^{{V_2}} \tilde{R}_{v_2} \\
\text{s.t.} \quad
& \tilde{R}_{v_1} \geq R_{\text{th}}, \\
& \mathbf{C}_{v_l} \succeq 0, \quad v_l \in \Phi _l, \quad l \in \{1,2\}, \\
&  \sum_{v=1}^{V} \operatorname{Tr}({\mathbf{E}}_{{m}}{\mathbf{C}}_{{v}}) \leq P_m, \qquad \forall m = 1,\dots,M, 
\end{align}
\end{subequations}
where ${{\tilde R}_{{v_l}}},{v_l} \in {\Phi _l},l \in \{ 1,2\}$,​​ is defined by applying the first-order Taylor series expansion to ${{R}_{{v_l}}}$:
\begin{align}
\tilde{R}_{v_l} &= B\, b_l \log_2\!\Bigg( \sum_{i=1}^{V_l} \mathrm{Tr}\big(\mathbf{A}_{v_l}\mathbf{C}_i\big) + \sigma^2 \Bigg) \nonumber\\
&\quad - B\, b_l \Bigg[ 
\log_2\!\Bigg( \sum_{\substack{i=1\\ i\ne v_l}}^{V_l} \mathrm{Tr}\big(\mathbf{A}_{v_l}\mathbf{C}_i^{(r)}\big) + \sigma^2 \Bigg) \nonumber\\
&\qquad\;
+ \sum_{\substack{i=1\\ i\ne v_l}}^{V_l}
\frac{\log_2 e \;\; \mathrm{Tr}\!\big(\mathbf{A}_{v_l}(\mathbf{C}_i - \mathbf{C}_i^{(r)})\big)}
{\displaystyle \sum_{\substack{i=1\\ i\ne v_l}}^{V_l}\mathrm{Tr}\big(\mathbf{A}_{v_l}\mathbf{C}_i^{(r)}\big) + \sigma^2}
\Bigg].
\end{align}

In (10), the variable \( \mathbf{C}_i^{(r)} \) denotes the value of \( \mathbf{C}_i \) at the \( r \)-th iteration.

At each iteration, Problem (9) is convex and can be solved using standard solvers such as CVX [23]. The corresponding procedure is outlined in Algorithm 1.

\begin{algorithm}[t]
\caption{Iterative Algorithm for Beamforming Design via the SCA Method}
\label{alg:SCA}
\begin{algorithmic}[1]

\State \textbf{Initialization:}
\State Set $r = 0$ ($r$ is the iteration number)
\State Initialize $\mathbf{C}_{v_l}^{(0)}$, $\forall v_l \in \Phi_l$, $l \in \{1,2\}$

\While{true}
    \State Update $r \gets r + 1$
    \State With given $\mathbf{C}_{v_l}^{(r)}$, solve problem (9) to obtain
    $\mathbf{C}_{v_l}^{(r+1)}$
    \State Update the beamforming vector $\mathbf{w}_{v_l}$ based on
    $\mathbf{C}_{v_l}^{(r+1)}$
    \If{the value of objective function (7) differs from the previous iteration
    by less than $\varepsilon$}
        \State \textbf{break}
    \EndIf
\EndWhile

\end{algorithmic}
\end{algorithm}

\subsubsection{Closed-form Solution for Beamforming Design}

In addition to the SCA-based method, a closed-form approach is developed in this section to solve the optimization problem in (7). This approach introduces a set of auxiliary variables, based on which the UAV beamforming vector is derived.

By introducing the auxiliary variables \(\boldsymbol{\alpha}_l = [\alpha_{l,1}, \ldots, \alpha_{l,V_l}] \), \( l \in \{1,2\} \), and applying the Lagrangian dual transform [24]-[25], problem~(7) can be equivalently reformulated as:
\begin{subequations}\label{eq:dual-transformed}
\begin{align}
\max_{\substack{
\boldsymbol{\alpha}_l,\, \mathbf{w}_{v_l} \\
l \in \{1,2\}
}} \quad 
& \sum_{v_2 = 1}^{V_2} 
\Big[
\log_2(1+\alpha_{2,v_2}) - \alpha_{2,v_2}
+ (1+\alpha_{2,v_2})\tfrac{\gamma_{v_2}}{1+\gamma_{v_2}}
\Big] \label{eq:dual-transformed-obj} \\
\text{s.t.} \quad 
& \log_2(1+\alpha_{1,v_1}) - \alpha_{1,v_1}
+ (1+\alpha_{1,v_1})\tfrac{\gamma_{v_1}}{1+\gamma_{v_1}}
\ge R_{\text{th}}, \ \ 
\label{eq:dual-transformed-constraint1} \\
& \sum_{v=1}^{V} \|\mathbf{w}_{v,m}\|^2 \le P_m,
\ \forall m=1,\dots,M,
\label{eq:dual-transformed-constraint2}
\end{align}
\end{subequations}
where the signal-to-interference-plus-noise ratio (SINR) of user \(v_l\) is given by:
\begin{equation}
\begin{aligned}
\gamma_{v_l} = \frac{|\mathbf{h}_{v_l}^H \mathbf{w}_{v_l}|^2}{\sum\limits_{\substack{i = 1 \\ i \ne v_l}}^{V_l} |\mathbf{h}_{v_l}^H \mathbf{w}_i|^2 + \sigma^2}, \quad v_l \in \Phi_l,\ l \in \{1,2\}.
\end{aligned}
\end{equation}

According to~[24], when $\mathbf{w}_{v_l}$ is fixed, 
the optimal value of $\alpha_{l,v_l}$ in~(11) is given by
\begin{equation}
\alpha_{l,v_l}^\star = \gamma_{v_l}.
\end{equation}

Given the multiple-ratio structure of problem (11), and by applying the quadratic transform [25]-[26], the objective function can be reformulated using a new set of auxiliary variables $\beta_{2,v_2}$, $v_2 \in \Phi_2$. Specifically, with the aid of (12), the objective function in (11) is reformulated as:
\begin{equation}
\begin{aligned}
f(\boldsymbol{\alpha}_2,\boldsymbol{\beta}_2,\mathbf{w}) =\; &
\sum_{v_2=1}^{V_2} \log_2(1+\alpha_{2,v_2})
- \sum_{v_2=1}^{V_2} \alpha_{2,v_2} \\
& + \sum_{v_2=1}^{V_2}
2\sqrt{1+\alpha_{2,v_2}}\,
\Re\!\left\{ \beta_{2,v_2}^*\,\mathbf{h}_{v_2}^H\mathbf{w}_{v_2} \right\} \\
& - \sum_{v_2=1}^{V_2} |\beta_{2,v_2}|^2
\left( \sum_{i=1}^{V_2} \left| \mathbf{h}_{v_2}^H \mathbf{w}_i \right|^2
+ \sigma^2 \right).
\end{aligned}
\end{equation}

The optimal value of the auxiliary variable $\beta_{2,v_2}$ is obtained by taking the partial derivative of \( f \) with respect to $\beta_{2,v_2}^*$, while keeping all other variables fixed. The resulting closed-form solution is given by
\begin{equation}
\beta_{2,v_2}^\star =
\sqrt{1+\alpha_{2,v_2}}\,
\frac{\mathbf{h}_{v_2}^H \mathbf{w}_{v_2}}
{\sum\limits_{i=1}^{V_2} |\mathbf{h}_{v_2}^H \mathbf{w}_i|^2 + \sigma^2}.
\end{equation}

After applying the quadratic transform and determining the optimal values for the auxiliary variables $\beta_{2,v_2}$, we proceed to solve for the optimal beamforming vectors \(\mathbf{w}_{v_2}\). With fixed values of $\alpha_{2,v_2}$ and $\beta_{2,v_2}$, the beamforming design subproblem becomes convex and admits a closed-form solution. Specifically, by taking the derivative of (14) with respect to $\mathbf{w}_{v_2}$ and incorporating the corresponding Lagrange multipliers associated with the power constraints, the optimal beamforming vector can be expressed as
\begin{equation}\label{eq:w_v2_star_block}
\mathbf{w}_{v_2}^\star =
\sqrt{1+\alpha_{2,v_2}}\,\beta_{2,v_2}
\Big(
\boldsymbol{\Lambda}
+ \sum_{i=1}^{V_2} |\beta_{2,i}|^2\,\mathbf{h}_i\mathbf{h}_i^H
\Big)^{-1}
\mathbf{h}_{v_2},
\end{equation}
where
\begin{equation}
\boldsymbol{\Lambda} = \mathrm{blkdiag}(\lambda_1 \mathbf{I}_{N_1}, \lambda_2 \mathbf{I}_{N_2}, \ldots, \lambda_M \mathbf{I}_{N_M}),
\end{equation}
is a block-diagonal matrix collecting the Lagrange multipliers $\{\lambda_m\}_{m=1}^M$ in a matrix form.

Similarly, for the emergency vehicles, the optimal beamforming vector $\mathbf{w}_{v_1}$ is given by
\begin{equation}\label{eq:17_modified}
\mathbf{w}_{v_1}^\star 
= 
\sqrt{1 + \alpha_{1,v_1}}\, \beta_{1,v_1}
\left(
\boldsymbol{\Lambda}
+ 
\sum_{i = 1}^{V_1} 
|\beta_{1,i}|^2 
\, \mathbf{h}_i \mathbf{h}_i^H
\right)^{-1}
\mathbf{h}_{v_1},
\end{equation}
where $\beta_{1,v_1}^\star=
\sqrt{1 + \alpha_{1,v_1}}
\frac{\mathbf{h}_{v_1}^H \mathbf{w}_{v_1}}
{\sum_{i = 1}^{V_1} \left| \mathbf{h}_{v_1}^H \mathbf{w}_{i} \right|^2 + \sigma^2}.$, and the Lagrange multipliers $\{\lambda_m\}_{m=1}^M$ must be appropriately chosen such that the UAVs' power constraints and the minimum rate requirement for emergency vehicles are satisfied. Accordingly, the optimal multipliers are obtained as

\begin{equation}
\boldsymbol{\Lambda} =
\arg\min_{\substack{\lambda_m \succeq 0 \\ m=1,\dots,M}}
 \sum_{v=1}^{V} \|\mathbf{w}_{v,m}\|^2 \le P_m.
\label{eq:Lambda_opt}
\end{equation}

The matrix $\boldsymbol{\Lambda}$, which contains the Lagrange multipliers of the power constraints, is adjusted to satisfy the minimum rate requirement of emergency vehicles.

To assist in computing the Lagrange multipliers $\{\lambda_m\}_{m=1}^M$, the following auxiliary matrices are introduced:
\begin{equation}
\begin{gathered}
\mathbf{A}_{v_l} = \sum_{i = 1}^{V_l} |\beta_{l,i}|^2 \mathbf{h}_i \mathbf{h}_i^H,
\qquad
\mathbf{B}_{v_l} = \mathbf{A}_{v_l} + \boldsymbol{\Lambda}. \\ 
\end{gathered}
\end{equation}

If the matrix $\mathbf{B}_{{v_l}}$ is invertible and the following conditions are satisfied, the corresponding beamforming vectors can be directly calculated.

\begin{align}
\sum_{v=1}^{V} \|\mathbf{w}_{v,m}^{(0)}\|^2 &= P_m, \nonumber\\
B b_1 \log_2\!\left(1 + \frac{|\mathbf{h}_{v_1}^H \mathbf{w}_{v_1}^{(0)}|^2}
{\sum_{i\ne v_1} |\mathbf{h}_{v_1}^H \mathbf{w}_i^{(0)}|^2 + \sigma^2}\right) &\ge R_{\text{th}},\quad v_1 \in \ \Phi_1,
\end{align}
where ${\mathbf{w}}_{v_l}^{(0)}$ denotes the initial value of the beamforming vector ${\mathbf{w}}_{v_l}$.
Otherwise, the following power constraint must be satisfied:
\begin{equation}
\sum_{v = 1}^{V} \operatorname{Tr}(\mathbf{w}_{v,m} \mathbf{w}_{v,m}^H) = P_m,  m=1,\dots,M.
\end{equation}

\begin{algorithm}[t]
\caption{Closed-form Beamforming Design Procedure}
\label{alg:CF}
\begin{algorithmic}[1]

\State \textbf{Initialization:}
\State Set iteration index $r = 0$
\State Initialize $\mathbf{w}_{v_l}^{(0)}$, $\forall v_l \in \Phi_l$, $l \in \{1,2\}$, such that
\State \hspace{1em} $\sum_{v=1}^{V} \|\mathbf{w}_{v,m}^{(0)}\|^2 = P_m$
\State \hspace{1em} $B b_1 \log_2\!\left(1 + \gamma_{1,v_1}^{(0)}\right) \ge R_{\text{th}},\ \forall v_1 \in \Phi_1$

\Repeat
\State Update iteration index: $r \gets r + 1$

\State \textbf{Auxiliary variable update:}
\For{$l \in \{1,2\}$, $v_l \in \Phi_l$}
\State \hspace{1em}
$\alpha_{l,v_l}^{(r)} =
\frac{|\mathbf{h}_{v_l}^H \mathbf{w}_{v_l}^{(r)}|^2}
{\sum_{i \ne v_l} |\mathbf{h}_{v_l}^H \mathbf{w}_i^{(r)}|^2 + \sigma^2}$
\State \hspace{1em}
$\beta_{l,v_l}^{(r)} =
\sqrt{1+\alpha_{l,v_l}^{(r)}}
\frac{\mathbf{h}_{v_l}^H \mathbf{w}_{v_l}^{(r)}}
{\sum_{i=1}^{V_l} |\mathbf{h}_{v_l}^H \mathbf{w}_i^{(r)}|^2 + \sigma^2}$
\EndFor

\State \textbf{Beamforming update:}
\For{$l \in \{1,2\}$}
\State \hspace{1em}
$\mathbf{Q}_l^{(r)} =
\boldsymbol{\Lambda}^{(r)}
+ \sum_{i=1}^{V_l} |\beta_{l,i}^{(r)}|^2 \mathbf{h}_i \mathbf{h}_i^H$
\For{$v_l \in \Phi_l$}
\State \hspace{2em}
$\mathbf{w}_{v_l}^{(r+1)} =
\sqrt{1+\alpha_{l,v_l}^{(r)}}\,\beta_{l,v_l}^{(r)}
\left(\mathbf{Q}_l^{(r)}\right)^{-1}
\mathbf{h}_{v_l}$
\EndFor
\EndFor

\State Compute the Lagrange multipliers $\{\lambda_m\}_{m=1}^{M}$ according to (24),
while ensuring that constraint (7b) is satisfied

\Until{the change in the objective value of problem (7) is less than $\varepsilon$}

\end{algorithmic}
\end{algorithm}

Given that the matrix \(\mathbf{B}_{v_l}\) is Hermitian, its eigendecomposition can be expressed as
\(\mathbf{B}_{v_l} = \mathbf{G}_{v_l} \boldsymbol{\Sigma}_{v_l} \mathbf{G}_{v_l}^H\),
where \(\mathbf{G}_{v_l}\) is a unitary matrix and $\boldsymbol{\Sigma}_{v_l}$ is a diagonal matrix containing the eigenvalues of \(\mathbf{B}_{v_l}\). Based on this decomposition, the power constraint in equation (22) can be equivalently reformulated as:
\begin{align}
\sum_{v_1=1}^{V_1} \operatorname{Tr}\Big[\big(\boldsymbol{\Sigma}_{v_1}+\lambda_m\mathbf{I}\big)^{-2}\mathbf{\Theta}_{v_1}\Big]
&\;\; \nonumber\\[4pt]
+\sum_{v_2=1}^{V_2} \operatorname{Tr}\Big[\big(\boldsymbol{\Sigma}_{v_2}+\lambda_m\mathbf{I}\big)^{-2}\mathbf{\Theta}_{v_2}\Big]
&\;=\; P_m.
\end{align}
where $\mathbf{\Theta}_{v_l} = \mathbf{G}_{v_l}^H \left( \sum_{i=1}^{V_l} (1 + \alpha_{l,i}) |\beta_{l,i}|^2 \, \mathbf{h}_{i} \mathbf{h}_{i}^H \right) \mathbf{G}_{v_l}$.
Using the basic properties of the trace operator, equation (23) can be further reduced to:

\begin{align}
\sum_{v_1=1}^{V_1}
\sum_{l=(m-1)L+1}^{mL}
\frac{(\mathbf{\Theta}_{v_1})_{l,l}}
     {\big((\boldsymbol{\Sigma}_{v_1})_{l,l}+\lambda_m\big)^2}
& \nonumber\\[4pt]
+\sum_{v_2=1}^{V_2}
\sum_{l=(m-1)L+1}^{mL}
\frac{(\mathbf{\Theta}_{v_2})_{l,l}}
     {\big((\boldsymbol{\Sigma}_{v_2})_{l,l}+\lambda_m\big)^2}
&= P_m,\qquad \forall m=1,\dots,M,
\end{align}
where the Lagrange multipliers $\{\lambda_m\}_{m=1}^M$ are computed via the bisection method [21], [27], and the resulting solution is accepted only if both constraints (7b) and (24) are satisfied. The overall procedure for deriving the closed-form beamforming solution is summarized in Algorithm 2. To obtain an initial point that respects the conditions in lines 4–5 of Algorithm 2, the following three-step procedure is employed:

\begin{itemize}
\item For every vehicle \(v\), set the initial beamforming vector as
\begin{equation}
\mathbf{w}_{v}^{(0)} = \rho_0 \frac{\mathbf{h}_v}{\|\mathbf{h}_v\|},\qquad 
\rho_0 = \sqrt{\frac{P_m}{V L}}.
\end{equation}

\item For each UAV \(m\), compute its total transmit power \(P_{\text{used}}^{(m)} = \sum_{v=1}^{V} \|\mathbf{w}_{v,m}^{(0)}\|^2\). If \(P_{\text{used}}^{(m)} > P_m\), scale down all beamforming vectors of that UAV by a factor slightly smaller than \(1\) (e.g., \(0.99\sqrt{P_m/P_{\text{used}}^{(m)}}\)) so that the constraint \(\sum_{v}\|\mathbf{w}_{v,m}^{(0)}\|^2 \le P_m\) is satisfied.

\item Compute the SINR and rate of each emergency vehicle using the current \(\mathbf{w}_{v_1}^{(0)}\). If the rate of any emergency vehicle is below the threshold \(R_{\mathrm{th}}\), multiply its beamforming vector by a factor slightly larger than \(1\) (e.g., \(1.2\)) and repeat the power scaling step. This boosting loop continues for a small number of iterations (e.g., \(10\)–\(15\)) until all emergency rates satisfy \(R_{\mathrm{th}}\).
\end{itemize}

  \begin{figure}[]
\centering
\begin{center}
\includegraphics[trim={0cm 0cm 0cm 0cm},clip, width=8.4cm, height=6.5cm]{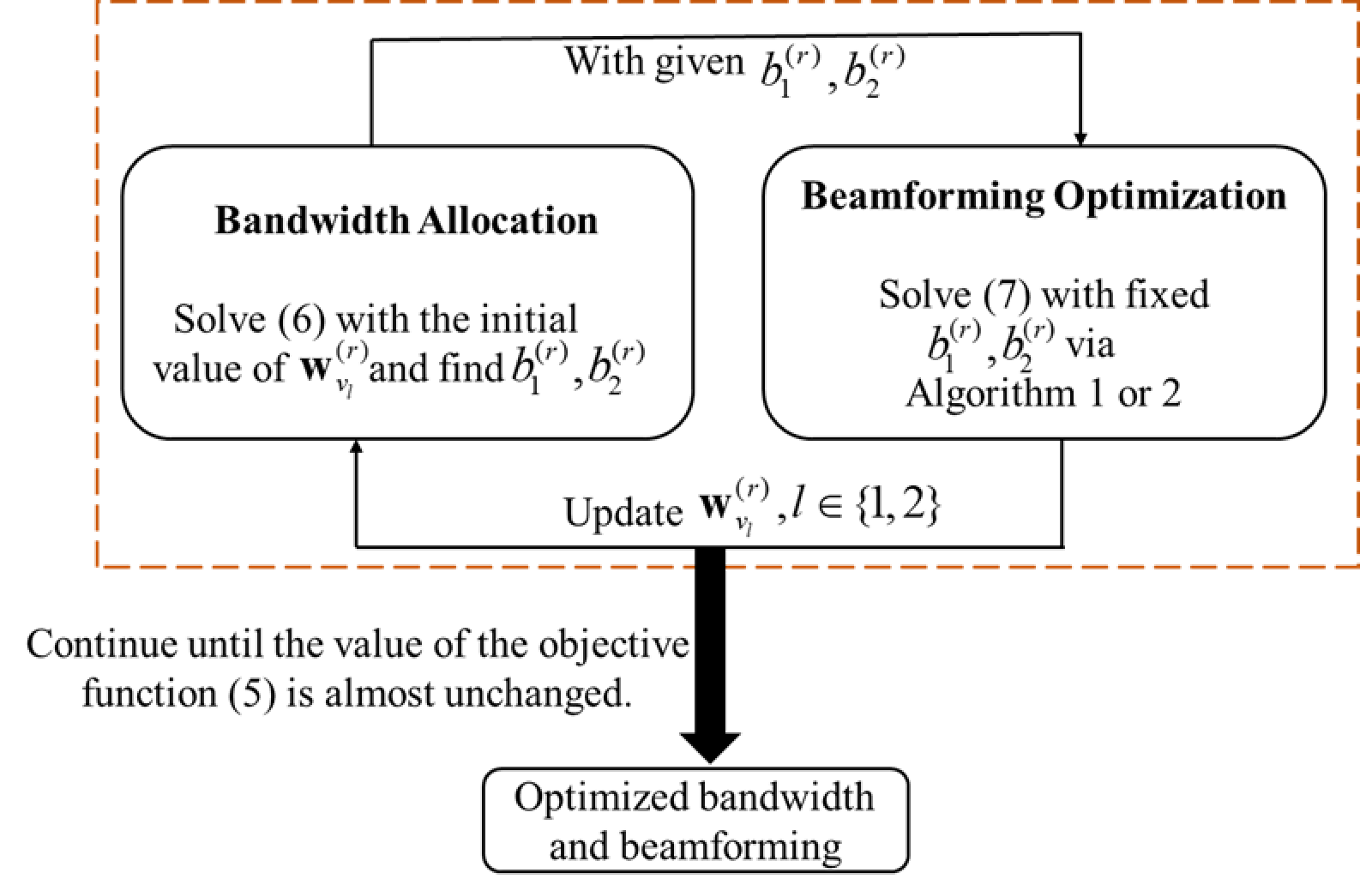}
\caption{Two-step iterative optimization method.}
\end{center}
\label{fig1}
\end{figure}

\subsection{Overall Algorithm and Computational Complexity Analysis}

\noindent To tackle the main optimization problem (5) using the block coordinate ascent (BCA) framework, as illustrated in Fig.~2, the original problem is decomposed into two subproblems that are solved iteratively. Subproblem~1 optimizes the bandwidth allocation for both regular and emergency vehicle groups while keeping the UAV beamforming vectors fixed. This subproblem corresponds to optimization problem (6), which is convex and can be solved using standard convex solvers such as CVX. Subproblem~2 updates the beamforming vectors based on the bandwidth values obtained in the previous step. Then, this subproblem can be solved either through the SCA method or by employing a closed‑form approach, as detailed in Section~III-B. This algorithm is convergent, as detailed in the Appendix A. The two subproblems are alternately solved until a sufficient improvement in the objective function is observed between successive iterations. The overall procedure is summarized in Algorithm~3.

The computational complexity of the SCA‑based approach is formulated as:
\begin{equation}
O\left(R_1 N_{\mathrm{IP}} (V (M L)^2)^{3.5} \log\frac{1}{\varepsilon} \right),
\end{equation}
where \(R_1\) denotes the number of iterations used in this method, as described in Algorithm~1, and \(N_{\mathrm{IP}}\) denotes the number of iterations required by Algorithm~3. The SDP in this method has \(n = V(ML)^2\) scalar variables (each \(\mathbf{C}_v\) is an \((ML)\times(ML)\) Hermitian matrix) and the interior‑point method per‑iteration complexity is \(O(n^{3.5})\) [21].

\begingroup
\AtBeginEnvironment{algorithm}{\color{black}\normalcolor\small}  
\color{black}
\begin{algorithm}[t]
\caption{Joint Bandwidth and Beamforming Optimization}
\label{alg:JointOpt}
\begin{algorithmic}[1]
\State \textbf{Initialization:} Set $r=0$, $b_1^{(0)}$, $b_2^{(0)}$.
\Repeat
    \State Fix beamforming, solve (6) to get $b_1^{(r)}, b_2^{(r)}$.
    \State Fix bandwidth, solve (7) using Alg.~1 or Alg.~2.
    \State $r \gets r + 1$.
\Until{change in objective (5) $< \varepsilon$}
\end{algorithmic}
\end{algorithm}
\endgroup

Additionally, the computational complexity of the closed‑form solution is given by [25]:
\begin{equation}
O\left(R_2 N_{\mathrm{IP}} V M (M L)^3 \log\frac{\lambda_u - \lambda_l}{\varepsilon} \right),
\end{equation}
where \(R_2\) is the required iterations to compute the beamforming vector based on Algorithm~2, and \(\lambda_u\) and \(\lambda_l\) represent the upper and lower bounds employed in the bisection search for the Lagrange multipliers \(\{\lambda_m\}_{m=1}^{M}\). In this method, inverting the \((ML)\times(ML)\) matrix in (16) costs \(O((ML)^3)\). This inversion is performed for each of the \(V\) users, giving \(O(V(ML)^3)\). Solving the per‑UAV power constraints via bisection on (24) introduces the extra factor \(M\).\\
As observed, the closed‑form solution exhibits a lower computational complexity compared to the SCA‑based method, although the SCA‑based method remains valuable as a general‑purpose framework that can readily accommodate additional constraints.

\section{Numerical Results}

In this section, we provide numerical results to assess the performance of the proposed optimization framework. The main simulation parameters used in the evaluations are summarized in Table~I, including the total available transmit power of the system. In the numerical analysis, four vehicles are considered, among which two correspond to emergency users. Moreover, the UAVs are assumed to be deployed with equal spacing along the road to ensure adequate coverage throughout the service area. The total system bandwidth is set to \(B=5\)~MHz, and the total transmit power \(P\) is equally shared among the UAVs such that \(P_m = P/M\). In addition, the minimum rate threshold for emergency vehicles is set to \( R_{\text{th}} = 32 \) Kbps.\footnote{This value is a typical data rate for safety message delivery in V2X applications [30].}

\begin{table}[]
\centering
\caption{Simulation Parameters}
\label{tab1}
\renewcommand{\arraystretch}{1.5}
\begin{tabular}{|c|c|}
\hline
\textbf{Parameter} & \textbf{Value} \\
\hline
\rule{0pt}{2.4ex} UAV flight altitude, \(z\) & 100\,m [28] \\
\hline
\rule{0pt}{2.4ex} Channel gain at 1\,m, \(d_0\) & \(-30\)\,dB [29] \\
\hline
\rule{0pt}{2.4ex} Receiver noise power at each vehicle, \(\sigma^2\) & \(-110\)\,dBm \\
\hline
\rule{0pt}{2.4ex} Total transmit power, \(P\) & 5\,W \\
\hline
\rule{0pt}{2.4ex} Length of road area, \(E_x\) & 5000\,m \\
\hline
\rule{0pt}{2.4ex} Width of road, \(E_y\) & 50\,m \\
\hline
\rule{0pt}{2.4ex} Iteration threshold, \(\varepsilon\) & \(10^{-4}\) \\
\hline
\rule{0pt}{2.4ex} Number of UAVs, \(M\) & 3 \\
\hline
\rule{0pt}{2.4ex} Number of antennas per UAV, \(L\) & 2 \\
\hline
\end{tabular}
\end{table}

\begin{figure}[htbp]
\centering
\begin{tikzpicture}
\begin{axis}[
    width=0.6\columnwidth,
    height=0.6*\columnwidth,
    grid=both,
    xlabel={Number of iterations},
    ylabel={Sum rate of regular vehicles (bps)},
    legend style={at={(0.02,0.4)}, anchor=north west, font=\footnotesize},
    tick label style={font=\small},
    label style={font=\small},
    xmin=0.5, xmax=20.5,
    xtick={1,5,10,15,20},
    scaled y ticks=base 10:-6,
    y tick scale label style={
        at={(axis description cs:0.02,1.02)},
        anchor=south west
    }
]

\addplot[red, mark=o, thick] table {JBB_CF_3.dat};
\addlegendentry{JBB-CF}

\addplot[blue, mark=square, thick] table {JBB_S_3.dat};
\addlegendentry{JBB-S}

\addplot[green!60!black, mark=triangle, thick] table {BF_only_3.dat};
\addlegendentry{BF[10]}

\addplot[black, mark=diamond, thick] table {BA_only_3.dat};
\addlegendentry{BA}

\end{axis}
\end{tikzpicture}
\captionsetup{justification=centering}
\caption{Convergence behavior of the proposed algorithms.}
\label{fig:convergence}
\end{figure}
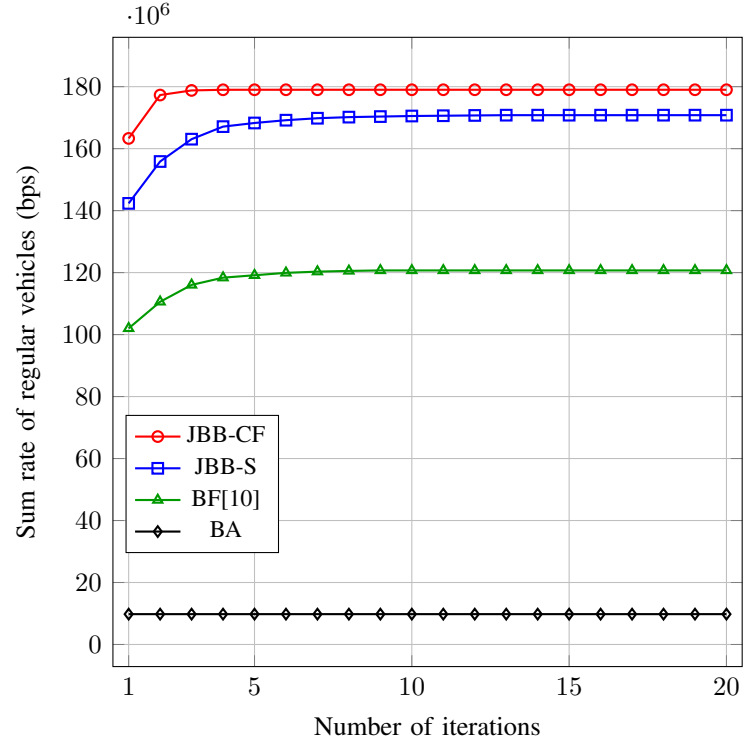


Fig.~3 illustrates the convergence of the sum rate of regular vehicles as a function of the number of iterations under four different resource allocation schemes:
\begin{itemize}
    \item JBB-CF: joint beamforming and bandwidth allocation using the closed-form algorithm,
    \item JBB-S: joint beamforming and bandwidth allocation using the SCA-based algorithm,  
    \item BF: beamforming optimization only, with equal bandwidth allocation~[10],
    \item BA: bandwidth allocation only, with fixed beamforming vectors.
\end{itemize}

As illustrated in this figure, as the number of iterations increases, the sum rate of the joint optimization schemes (JBB-CF and JBB-S) improves and gradually converges. The JBB-CF method achieves slightly higher rates than JBB-S, indicating the advantage of the closed‑form solution. The BF method achieves a higher rate than BA, because it has up to $2MLV$ degrees of freedom (real variables) for the beamforming vectors, whereas BA only has two degrees of freedom ($b_1$ and $b_2$). The BA curve remains constant because with fixed beamforming vectors, the bandwidth allocation problem is convex and its optimal solution is obtained directly.

\begin{figure}[]
\centering
\begin{tikzpicture}
\begin{axis}[
    width=0.6*\columnwidth,
    height=0.6*\columnwidth,
    grid=both,
    xmin=4.5, xmax=25.5,
    ymin=0, ymax=2.1e8,
    xlabel={Transmit power $P$ (W)},
    ylabel={Sum rate of regular vehicles (bps)},
    legend style={at={(0.02,0.35)}, anchor=west, font=\footnotesize},
    tick label style={font=\small},
    label style={font=\small},
    scaled y ticks=base 10:-6,
    y tick scale label style={
        at={(axis description cs:0.02,1.02)},
        anchor=south west
    }
]

\addplot[
    color={rgb,1:red,0.00;green,0.45;blue,0.74},
    mark=o,
    thick
] table[x index=0,y index=1] {sumrate_vs_power.dat};
\addlegendentry{JBB-CF}

\addplot[
    color={rgb,1:red,0.85;green,0.33;blue,0.10},
    mark=square,
    thick
] table[x index=0,y index=2] {sumrate_vs_power.dat};
\addlegendentry{JBB-S}

\addplot[
    color={rgb,1:red,0.47;green,0.67;blue,0.19},
    mark=triangle,
    thick
] table[x index=0,y index=3] {sumrate_vs_power.dat};
\addlegendentry{BF [10]}

\addplot[
    color=black,
    mark=diamond,
    thick
] table[x index=0,y index=4] {sumrate_vs_power.dat};
\addlegendentry{BA}

\end{axis}
\end{tikzpicture}
\captionsetup{justification=centering}
\caption{Sum rate of regular vehicles versus total transmit power of UAVs.}
\label{fig:sumrate_power}
\end{figure}
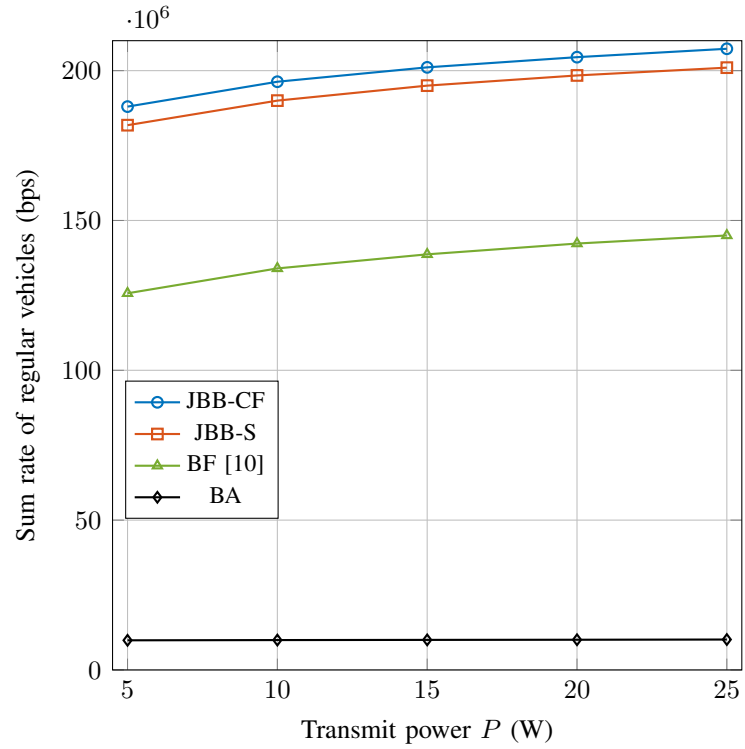

Fig. 4 illustrates the sum rate of regular vehicles versus the total transmit power of UAVs. As the transmit power increases, all schemes exhibit an upward trend due to improved received signal quality. The joint schemes consistently achieve higher sum rates than the BF and BA baselines, thereby validating the benefit of jointly optimizing both resources.

\begin{figure}[]
\centering
\begin{tikzpicture}
\begin{axis}[
    width=0.6\columnwidth,
    height=0.6*\columnwidth,
    grid=both,
    xmin=4.5, xmax=25.5,
    xlabel={Total transmit power $P$ (W)},
    ylabel={Sum rate of regular vehicles (bps)},
    legend style={at={(0.02,0.55)}, anchor=west, font=\footnotesize},
    tick label style={font=\small},
    label style={font=\small},
    scaled y ticks=base 10:-6,
    y tick scale label style={
        at={(axis description cs:0.02,1.02)},
        anchor=south west
    }
]

\addplot[blue, mark=o, thick] table {M3_N2.dat};
\addlegendentry{$M=3,\; N=2$}

\addplot[red, mark=square, thick] table {M2_N3.dat};
\addlegendentry{$M=2,\; N=3$}

\addplot[green!60!black, mark=triangle, thick] table {M2_N2.dat};
\addlegendentry{$M=2,\; N=2$}

\addplot[purple, mark=diamond, thick] table {M1_N6.dat};
\addlegendentry{$M=1,\; N=6$}

\addplot[black, mark=x, thick] table {M1_N2.dat};
\addlegendentry{$M=1,\; N=2$}

\end{axis}
\end{tikzpicture}
\captionsetup{justification=centering}
\caption{Sum rate of regular vehicles versus total transmit power of UAVs for various UAV and antenna configurations.}
\label{fig1}
\end{figure}
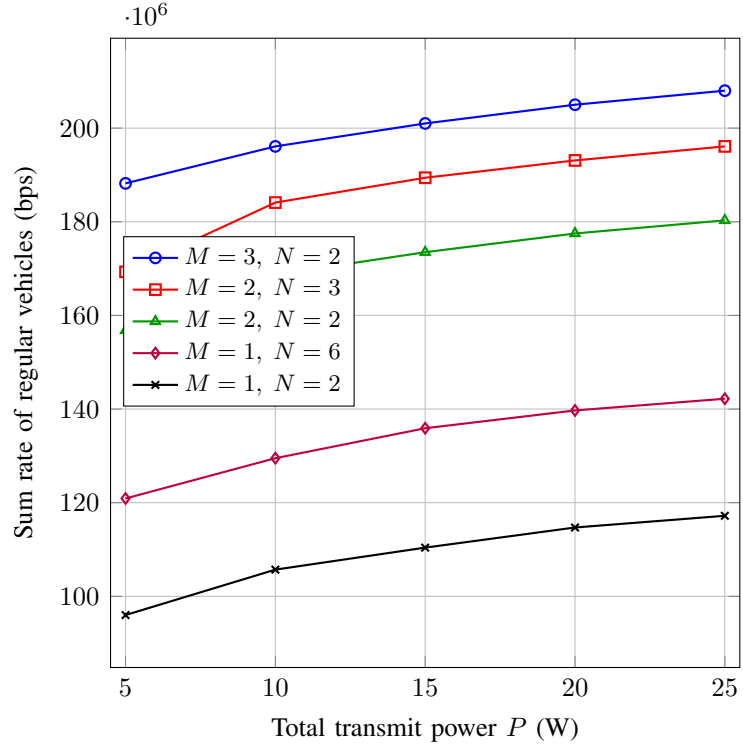

Fig. 5 illustrates the sum rate performance of regular vehicles versus the total transmit power for different configurations of the number of UAVs ($M$) and antennas per UAV ($L$) under the JBB-CF scheme. As expected, increasing the transmit power improves the achievable sum rate in all configurations due to the enhanced received signal quality. Among the considered setups, the configuration with $M=3$ UAVs and $L=2$ antennas yields the highest sum rate, followed by $M=2$, $L=3$ and $M=2$, $L=2$. The cases with a single UAV, i.e., $M=1$, generally result in lower rates, particularly when equipped with fewer antennas. These results highlight how the system performance can be influenced by the number of UAVs and their antenna capabilities in the given scenario.

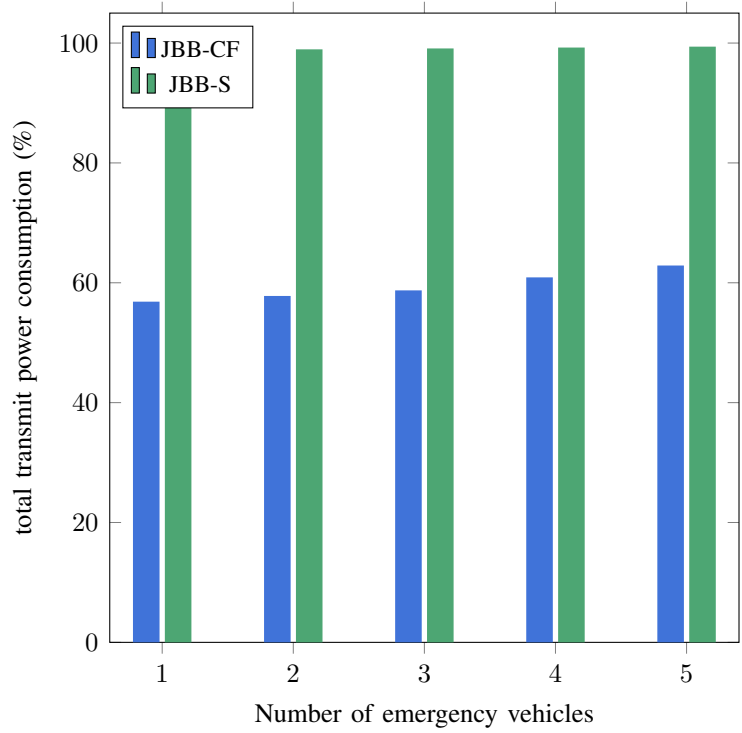
\begin{figure}[]
\centering
\begin{tikzpicture}
\begin{axis}[
    width=0.6\columnwidth,
    height=0.6\columnwidth,
    ybar,
    bar width=10pt,
    ymin=0, ymax=105,
    xlabel={Number of emergency vehicles},
    ylabel={total transmit power consumption (\%)},
    xtick={1,2,3,4,5},
    legend style={at={(0.02,0.98)}, anchor=north west, font=\footnotesize},
    tick label style={font=\small},
    label style={font=\small},
]

\addplot[
    fill={rgb,1:red,0.25;green,0.45;blue,0.85},
    draw=none
] table[x index=0, y index=1] {JBB_CF_power.dat};
\addlegendentry{JBB-CF}

\addplot[
    fill={rgb,1:red,0.30;green,0.65;blue,0.45},
    draw=none
] table[x index=0, y index=1] {JBB_S_power.dat};
\addlegendentry{JBB-S}

\end{axis}
\end{tikzpicture}
\captionsetup{justification=centering}
\caption{Total transmit power consumption versus the number of emergency vehicles.}
\label{fig:power_ev}
\end{figure}

Fig. 6 illustrates the percentage of the total transmit power allocated to the UAVs as a function of the number of emergency vehicles. As observed, increasing the number of emergency users requires the system to allocate more transmit power to ensure that their minimum rate requirement is satisfied, as imposed by constraint (7b). Consequently, to guarantee the QoS of emergency vehicles while maximizing the rate of regular vehicles, a higher amount of transmit power is consumed. This leads to an overall increase in total transmit power consumption for both schemes as the number of emergency vehicles grows. Nevertheless, the JBB-CF method consistently exhibits a noticeably lower total transmit power consumption compared with the JBB-S approach.

\begin{figure}[]
\centering
\begin{tikzpicture}
\begin{axis}[
    width=0.65\columnwidth,
    height=0.6*\columnwidth,
    grid=both,
    xlabel={Total Bandwidth (MHz)},
    ylabel={Sum rate of regular vehicles (bps)},
    legend style={at={(0.02,0.98)}, anchor=north west, font=\footnotesize},
    tick label style={font=\small},
    label style={font=\small},
    xmin=1.8, xmax=10.2,
    scaled y ticks=base 10:-6,
    y tick scale label style={
        at={(axis description cs:0.02,1.02)},
        anchor=south west
    }
]

\addplot[blue, mark=o, thick] table {JBB_CF.dat};
\addlegendentry{JBB-CF}

\addplot[red, mark=square, thick] table {JBB_S.dat};
\addlegendentry{JBB-S}

\addplot[green!60!black, mark=triangle, thick] table {BF10.dat};
\addlegendentry{BF [10]}

\addplot[black, mark=diamond, thick] table {BA.dat};
\addlegendentry{BA}

\end{axis}
\end{tikzpicture}
\captionsetup{justification=centering}
\caption{Sum rate of regular vehicles versus total available bandwidth.}
\label{fig:sumrate_bandwidth}
\end{figure}
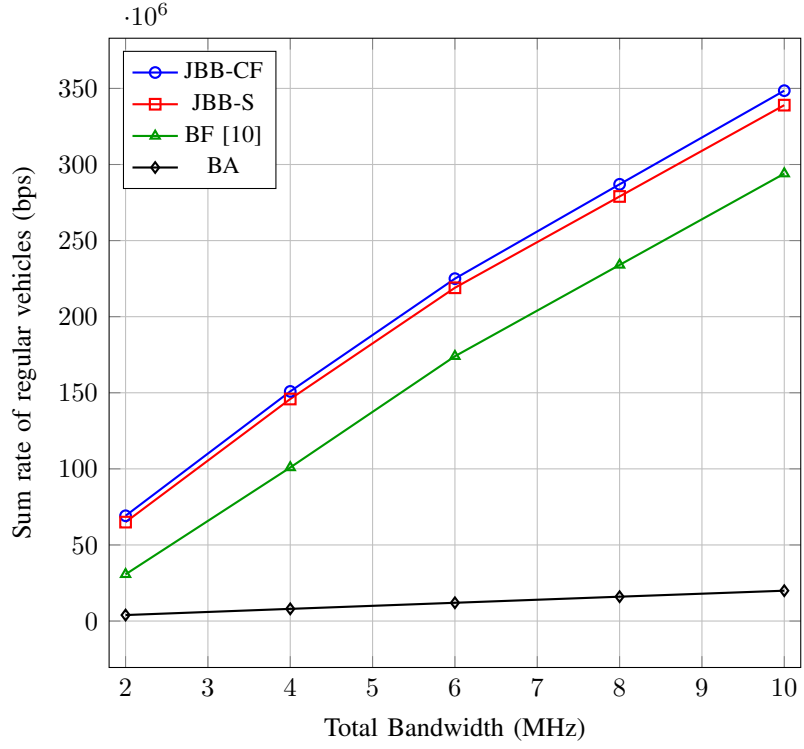

Fig. 7 illustrates the sum rate of regular vehicles as a function of the total available bandwidth. As the bandwidth increases, all schemes exhibit an upward trend, reflecting the direct relationship between bandwidth and achievable data rate. The JBB-CF scheme achieves better performance compared to the JBB-S scheme, and both joint designs significantly outperform the BF and BA schemes.

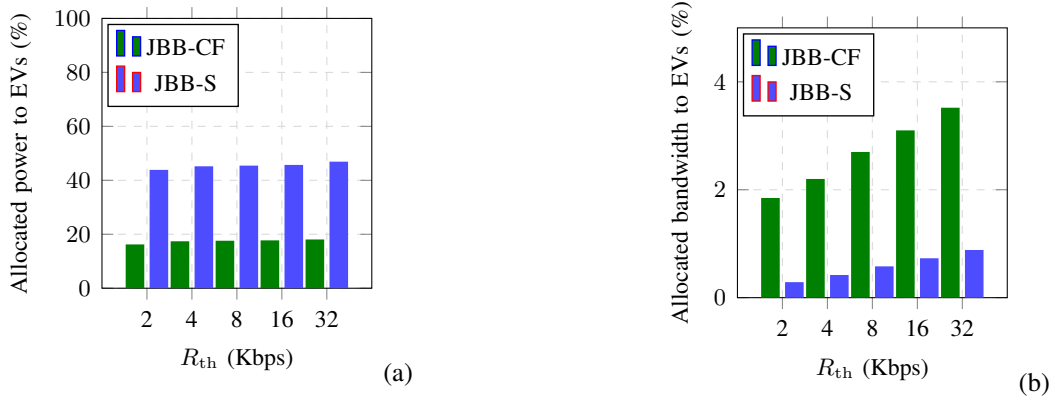
\begin{figure}[]
\centering

\centering
\begin{minipage}{0.48\textwidth}
\centering
\begin{tikzpicture}
\begin{axis}[
    ybar,
    bar width=7pt,
    width=0.65\linewidth,   
    height=0.65\linewidth,  
    enlarge x limits=0.25,
    ylabel={Allocated power to EVs (\%)},
    xlabel={$R_{\mathrm{th}}$ (Kbps)}, 
    xmode=log,
    log basis x=2,
    xtick={2,4,8,16,32},
    xticklabels={2,4,8,16,32},
    xmin=2, xmax=32,
    ymin=0, ymax=100,
    grid=major,
    grid style={dashed,gray!30},
    legend style={
        at={(0.02,0.98)},
        anchor=north west,
        font=\footnotesize,
        draw=black,
        line width=0.5pt
    },
    tick label style={font=\footnotesize},
    label style={font=\footnotesize},
]
\addplot+[fill=green!50!black, draw=none]
table[x index=0, y index=1] {Power_JBB_CF.dat};
\addplot+[fill=blue!70, draw=none]
table[x index=0, y index=1] {Power_JBB_S.dat};
\legend{JBB-CF, JBB-S}
\end{axis}
\end{tikzpicture}
\vspace{0.1cm}
{\small (a)}
\end{minipage}
\hfill
\begin{minipage}{0.48\textwidth}
\centering
\begin{tikzpicture}
\begin{axis}[
    ybar,
    bar width=7pt,
    width=0.65\linewidth,
    height=0.65\linewidth,
    enlarge x limits=0.25,
    ylabel={Allocated bandwidth to EVs (\%)},
    xlabel={$R_{\mathrm{th}}$ (Kbps)},
    xmode=log,
    log basis x=2,
    xtick={2,4,8,16,32},
    xticklabels={2,4,8,16,32},
    xmin=2, xmax=32,
    ymin=0, ymax=5,
    grid=major,
    grid style={dashed,gray!30},
    legend style={
        at={(0.02,0.98)},
        anchor=north west,
        font=\footnotesize,
        draw=black,
        line width=0.5pt
    },
    tick label style={font=\footnotesize},
    label style={font=\footnotesize},
]
\addplot+[fill=green!50!black, draw=none]
table[x index=0, y index=1] {BW_JBB_CF.dat};
\addplot+[fill=blue!70, draw=none]
table[x index=0, y index=1] {BW_JBB_S.dat};
\legend{JBB-CF, JBB-S}
\end{axis}
\end{tikzpicture}
\vspace{0.1cm}
{\small (b)}
\end{minipage}
\caption{Resource allocation to emergency vehicles versus the threshold $R_{\mathrm{th}}$: 
(a) power allocation percentage, (b) bandwidth allocation percentage.}
\label{fig:Rth_two_plots}
\end{figure}

Fig.~8 illustrates the percentage of power and bandwidth allocated to emergency vehicles as a function of the threshold $R_{th}$. As $R_{th}$ increases, the allocation of both resources grows in both the JBB-CF and JBB-S schemes. It can be observed that the JBB-CF scheme consumes less transmit power compared to JBB-S, while allocating a slightly higher portion of the total bandwidth to emergency vehicles. Moreover, in both schemes, the bandwidth assigned to emergency vehicles remains below 5\% of the overall system bandwidth.

\section{Conclusion}

This paper has investigated a UAV-assisted vehicular communication system designed to address diverse service requirements by categorizing vehicles into emergency and regular groups. By jointly optimizing bandwidth allocation and distributed beamforming, the proposed framework improves the sum data rates of regular vehicles while satisfying the minimum rate constraints of emergency users. 

Simulation results show that the joint optimization methods, including the closed-form (JBB-CF) and iterative SCA-based (JBB-S) algorithms, achieve notably better performance compared with approaches that optimize only beamforming or bandwidth allocation. The JBB-CF method attains higher sum rates than the JBB-S algorithm across the considered numerical evaluations. Furthermore, system performance increases with higher UAV transmit power, as well as with a larger number of UAVs and antennas per UAV, highlighting the benefit of deploying multi-antenna UAVs. Increasing the available bandwidth also raises the achievable data rates, showing the impact of resource allocation on system behavior. 
The numerical results further illustrate that the power and bandwidth assigned to emergency vehicles increase as their minimum rate requirement becomes more demanding. Between the two joint schemes, JBB-CF generally consumes less transmit power compared with JBB-S. These observations indicate that JBB-CF provides competitive performance, offering higher sum rates as well as  reduced transmit power consumption, and benefiting from lower complexity.

As possible next steps, incorporating UAV trajectory optimization and robust beamforming designs under imperfect CSI are promising directions for future work.

\section*{Acknowledgment}

The authors would like to thank the ICT Research Institute (Iran Telecommunication Research Center--ITRC) for its valuable support of this research.

\section*{APPENDIX A}
In this section, we provide the convergence proof for the proposed alternating optimization framework. The objective function for problem (5) is formulated as $f(b_2^{(r)},\mathbf{W}_2^{(r)})$ during the $r$-th iteration, where the matrix $\mathbf{W}_2^{(r)}$ collects the beamforming vectors of regular vehicles:
\begin{equation}
\mathbf{W}_2^{(r)} = [\mathbf{w}_{1}^{(r)}, \mathbf{w}_{2}^{(r)}, \ldots, \mathbf{w}_{V_2}^{(r)}].
\end{equation}
 \\At the optimal bandwidth $b_2^{(r+1)}$ obtained from solving problem (6), the following relation holds in step 4 of Algorithm 3:
\begin{equation}
f(b_2^{(r)},\mathbf{W}_2^{(r)}) \le f(b_2^{(r+1)},\mathbf{W}_2^{(r)}).
\end{equation} \\For regular vehicles, let $\boldsymbol{\Gamma}^{(r)} = \{\mathbf{C}_{v_2}^{(r)}\}_{v_2=1}^{V_2}$ denote the set of covariance matrices, with $\mathbf{C}_{v_2}^{(r)} = \mathbf{w}_{v_2}^{(r)}(\mathbf{w}_{v_2}^{(r)})^H$. The SCA‑based beamforming update (Algorithm 1) defines $\tilde f(b_2^{(r+1)},\boldsymbol{\Gamma})$ as the objective of the relaxed SDP (8), and $\tilde f^{L}(b_2^{(r+1)},\boldsymbol{\Gamma};\boldsymbol{\Gamma}^{(r)})$ as its linearized surrogate (10). The following relationships are then derived:
\begin{equation}
\begin{array}{l}
f(b_2^{(r+1)},\mathbf{W}_2^{(r)})\mathop  = \limits^a \tilde f(b_2^{(r+1)},\boldsymbol{\Gamma}^{(r)})\\[0.5 cm]
\mathop  = \limits^b \tilde f^{L}(b_2^{(r+1)},\boldsymbol{\Gamma}^{(r)};\boldsymbol{\Gamma}^{(r)})\mathop  \le \limits^c \tilde f^{L}(b_2^{(r+1)},\boldsymbol{\Gamma}^{(r+1)};\boldsymbol{\Gamma}^{(r)})\\[0.5 cm]
\mathop  \le \limits^d \tilde f(b_2^{(r+1)},\boldsymbol{\Gamma}^{(r+1)}) = f(b_2^{(r+1)},\mathbf{W}_2^{(r+1)})
\end{array}
\end{equation}
where (a) holds because each covariance matrix $\mathbf{C}_{v_2}^{(r)} = \mathbf{w}_{v_2}^{(r)}(\mathbf{w}_{v_2}^{(r)})^H$ represents the same Hermitian beamforming design as $\mathbf{W}_2^{(r)}$, and thus the optimization with respect to $\mathbf{W}_2^{(r)}$ can equivalently be expressed in terms of $\boldsymbol{\Gamma}^{(r)}$ [21]; (b) follows from the tightness of the Taylor approximation in (10) at $\boldsymbol{\Gamma}^{(r)}$ [31]; (c) is valid because problem (9) is solved optimally to obtain $\boldsymbol{\Gamma}^{(r+1)}$; and (d) arises from the fact that the surrogate $\tilde f^{L}$ is a lower bound of $\tilde f$. 

For the closed-form beamforming solution (Algorithm 2), the monotonicity is guaranteed by the properties of the Lagrangian dual and quadratic transforms [24]--[26], which ensure that the updates (13), (15), (16) and (18) never decrease the objective. \\Therefore, combining the two steps, we conclude:
\begin{equation}
f(b_2^{(r)},\mathbf{W}_2^{(r)}) \le f(b_2^{(r+1)},\mathbf{W}_2^{(r+1)}).
\end{equation}

Thus, after each complete iteration of Algorithm 3, the objective function in problem (5) is monotonically non‑decreasing. Since the objective is bounded above by the finite transmit power and bandwidth, convergence is guaranteed.

{}

\end{document}